\documentclass[twocolumn, showpacs]{revtex4-1}

\usepackage{amsmath}
\usepackage{amssymb}

\usepackage{graphicx}

\usepackage{color} 

\usepackage{microtype}

\usepackage{pgf,booktabs}

\def\ER{Erd\H{o}s-R\'enyi }
\def\BA{Barab\'{a}si-Albert }

\begin{document}

\title{Discovering universal statistical laws of complex networks}

\author{Stefano \surname{Cardanobile}}
\affiliation{Bernstein Center Freiburg \& Faculty of Biology, Albert-Ludwig University, 79104 Freiburg, Germany}

\author{Volker \surname{Pernice}}
\affiliation{Bernstein Center Freiburg \& Faculty of Biology, Albert-Ludwig University, 79104 Freiburg, Germany}

\author{Moritz \surname{Deger}}
\affiliation{Bernstein Center Freiburg \& Faculty of Biology, Albert-Ludwig University, 79104 Freiburg, Germany}

\author{Stefan \surname{Rotter}}
\altaffiliation{Electronic address: stefan.rotter@biologie.uni-freiburg.de}
\affiliation{Bernstein Center Freiburg \& Faculty of Biology, Albert-Ludwig University, 79104 Freiburg, Germany}

\date{\today}


\begin{abstract}
Different network models have been suggested for the topology underlying complex interactions in natural systems. 
These models are aimed at replicating specific statistical features encountered in real-world networks. 
However, it is rarely considered to which degree the results obtained for one 
particular network class can be extrapolated to real-world networks. 
We address this issue by comparing different {classical and more recently developed} 
network models with respect to 
their generalisation power, which we identify with large structural 
variability and absence of constraints imposed by the construction scheme.
{
After having identified the most variable networks, we address the issue
of which constraints are common to all network classes and are thus 
suitable candidates for being generic statistical laws of complex networks.
}
In fact, we find that generic, {not model-related} dependencies between 
different network characteristics do exist. 
This allows, for instance, to infer global features from local ones using regression models trained 
on networks with high generalisation power. 
{
Our results confirm and extend previous findings regarding the synchronisation
properties of neural networks. 
Our method seems especially relevant for large networks, which are difficult to map completely, like the neural networks in the brain. 
The structure of such large networks cannot be fully sampled with the present
technology. Our approach provides a method to estimate
global properties of under-sampled networks with good approximation.
Finally, we demonstrate on three different data sets 
(\emph{C.\ elegans}' neuronal network, \emph{R.\ prowazekii}'s metabolic
network, and a network of synonyms extracted from Roget's Thesaurus)
that real-world networks have statistical relations compatible with those 
obtained using regression models.
}
\end{abstract}

\maketitle


\section*{Introduction}
The development of models for the topology underlying complex interactions in 
natural systems
has attracted much attention in recent research~\citep{Newman2003a,Barabasi2004,Milo2004}.
Since statistical features of the structure of such systems are known to
exert strong influence on their dynamics~\citep{Arenas2008,Kitsak2010,Rubinov2010},
these models are commonly defined in a stochastic framework.
Indeed, in many cases parametric families of network models exist that can 
replicate specific statistics observed in real {networks} and also explain how 
these statistics arise.
Classical examples are the emergence of a giant connected component
in percolation phenomena~\citep{Erdos1959},
and the power-law degree distributions 
observed in real-world networks~\citep{Barabasi1999}. 
Dynamical systems on networks have recently received much attention.
The influence of certain structural features on dynamical 
properties, 
like synchronizability~\citep{Watts1998,Grabow2010a} and
controllability~\citep{Albert2000,Liu2011}
has been analyzed with the help of particular {network} models.
This fact calls for an evaluation of the efficiency of 
existing network models in sampling the space of real-world networks. 
In fact, it is unlikely that a small number of standard models can reproduce the
variability of networks observed in nature, 
but this problem is rarely addressed in literature. 
To circumvent this problem we base our analysis on several different network models to avoid singular relations that hold only for specific cases. Remaining relations among different structural network features can then with much greater certainty be assumed to hold generally. 
In particular, we take advantage of two recently developed 
advanced network models, multifractal networks~\citep{Palla2010,Palla2011}
and equilibrium random networks~\citep{Chung2002}. 
These new classes encompass networks of greater structural diversity in the
statistical ensemble than, for example, \ER graphs or small-world networks.

As a first main result, we conclude that multifractal networks and equilibrium random networks 
are the most variable ones with respect to the generated entropy. 
They present a good sampling basis, 
as only weak correlations between different graph properties 
are imposed by their construction principle.

The issue of whether global, in particular spectral, properties of networks
are predictable from local statistical properties has been debated in the scientific 
community
with both negative~\citep{Atay2006} and positive~\citep{Zhan2009} results.
Our second main result is that global network properties, also of spectral nature, 
are indeed statistically linked to network 
properties on a local level, and that these relations are also relevant for 
real-world networks.
This is achieved using multivariate linear regression on an 
appropriate set of regressors among the local features.

In particular, we study three different networks:
a synonym network extracted from Roget's Thesaurus~\citep{Pajek2006}, 
the metabolic network of the bacterium \emph{R.\ prowazekii}~\citep{Jeong2000},
and the neuronal network of the nematode \emph{C.\ elegans}~\citep{Varshney2011b}.
We find that the dependencies between certain features follow
the same law for network models and for real data, thus justifying
our approach.

Our third main result concerns one specific relation that was in fact detected 
with our new method: we demonstrate that the synchronization index, 
a quantity introduced to assess the inertia to 
synchronization of complex networks~\citep{Atay2006}, depends very strongly 
on the variance of the in-degree, a fact that may be of special interest for scientists studying network synchronization~\citep{Grabow2010a}.

\section*{Models}

\begin{table}
\caption{Symbols and concepts}\label{tab:concepts}
\begin{center}
\scriptsize
\begin{tabular}{lll}
\toprule
Symbol              & Description               \\
\midrule 
$\mathrm{mean}(M)$       & Complex number: mean of the set $M$    \\
$\mathrm{var}(M)$   & Positive real number: variance of the set $M$    \\
$\mathrm{std}(M)$   & Positive real number:\\
& standard deviation of the set $M$    \\
$\mathrm{corr}(P)$   & Real number in $[-1,1]$:\\& Pearson correlation coefficient
of pairs $P$    \\
$\mathrm{clust}( v)$   & Real number in $[0,1]$: Fraction of undirected\\
& triangles between neighbors of $ v$\\
$\mathrm{shell}^\pm( v)$   
& Positive integer:
In or out-shell of node $ v$
\\
\midrule
$A=(a_{ij})$                   & Matrix: adjacency matrix of a graph:\\
& $a_{ij}=1$ iff link $j\rightarrow i$ exists, 
otherwise 0\\
$\mathrm{Tr}(A)$        & Complex, number:
 trace of the matrix $A$ \\
$L=(\ell_{ij})$                   & Matrix: Laplace matrix of a graph \\
$ V$                   & Set: node set of a graph \\
$ E$                   & Set: edge set of a graph \\
$\Gamma^\pm( v)$                   
& Sets of nodes: 
nodes targeting to or targeted by $v$ \\
$\mathrm{deg}^\pm(v)$ & 
Integers: cardinality of $\Gamma^\pm( v)$\\
\midrule
BA          & \BA network \\
ER          & \ER network\\
EQR         & Equilibrium random networks\\
MF(n,k)     & Multifractal network class:\\
& $n$ initial squares, $k$ iterations\\
WS          & Watts-Strogatz network
\\
\bottomrule\\
\end{tabular}
\end{center}
\end{table}

Each of the network models (for a list of the networks considered here see 
Table~\ref{tab:concepts}) is defined by a set of parameters;
the rationale of the comparison is to first draw a random set of graph parameters, 
then draw a specific realization using these parameters,
and finally analyze the structural properties of the graph.
The parameters of most network models we analyze have to be
chosen in a bounded set. It is therefore a natural choice to randomize the
parameters using uniform (real or integer-valued) distributions.
We will refer to this algorithm as to the 
doubly stochastic generation process.
We kept the average connectivity (i.e.\ the expected fraction of
realized edges out of all possible edges) fixed for all network models. 
In our study we used the value 0.1 throughout.
This value generally resulted in relatively sparse networks with a 
large connected component. 
We concentrated our attention on directed networks, and, if necessary, 
we extended the original definitions to directed versions. 
For each realization of a network, 
we extract a {feature vector} $f^{c_n}_i$ of commonly used statistical descriptors, 
see Table~\ref{tab:measures}. The apex $c_n$ indicates the 
$n$\textsuperscript{th} instance of the network class $c$, the index $i$ indicates the feature.

\begin{table}
\caption{{Statistical descriptors
(thematic ordering as in figures)}}\label{tab:measures}
\begin{center}
\scriptsize
\begin{tabular}{lll}
\toprule
Symbol    & Complete Name         & Description   \\
Local Descriptors\\
\midrule
CCM               & Mean clustering       &
$\mathrm{mean}\left( \{ \mathrm{clust} ( v) \} \right)$    \\
CCV               & Clustering variance      &
$\mathrm{var}\left( \{ \mathrm{clust} ( v) \}   
\right) $  \\
IDV               & Variance of in-degrees       &
$\mathrm{var}\left( \{ \mathrm{deg}^+( v)  \} \right)$ \\
IOD               & In-out correlation       &
$\mathrm{corr}\left( \{ (\mathrm{deg}^+( v),\mathrm{deg}^-( v))  \} \right)$   \\
ODV               & Variance of out-degrees       &
$\mathrm{var}\left( \{ \mathrm{deg}^-( v) : 
 v \in  V \} \right)$ \\
\midrule
IPIC        & In-mean-in correlation: &\\&
\multicolumn{2}{l}{
$\mathrm{corr}
\{
\left( \deg^+( v), 
\mathrm{mean}\left( \{ \deg^+( v') : v' \in \Gamma^-( v) \}\right) \right)
\}$}
\\
IPOC        & In-mean-out-correlation: &\\&
\multicolumn{2}{l}{
$\mathrm{corr}
\{
\left( \deg^+( v), 
\mathrm{mean} \left( \{ \deg^-( v') : v' \in \Gamma^-( v) \} \right)\right)
\}$}
\\
OPIC        & Out-mean-in-correlation: &\\&
\multicolumn{2}{l}{
$\mathrm{corr}
\{
\left( \deg^-( v), 
\mathrm{mean} \left(\{ \deg^+( v') : v' \in \Gamma^-( v) \} \right)\right)
\}$}
\\
OPOC        & Out-mean-out-correlation: &\\&
\multicolumn{2}{l}{
$\mathrm{corr}
\{
\left( \deg^-( v), 
\mathrm{mean}\left( \{ \deg^-( v') : v' \in \Gamma^-( v) \} \right)\right)
\}$}
\\
\midrule
FRC               & Fraction of\\& recurrent connections      &
$\frac{\sum_{ij} a_{ij}a_{ji}}{\sum_{i,j}{a_{ij}}}$    \\
Global Descriptors\\
SR              & Spectral radius       &
$\max \{ |\lambda|: \lambda \in \sigma(A)\}$     \\
NTR             & Normalized trace      &
$\mathrm{mean}\left( \sigma(A)\right) $     \\
VEV             & Variance of eigenvalues       &
$\mathrm{var}\left( \sigma(A)\right)$      \\
SI~\cite{Atay2006}              & Synchronization index & 
$\max \{ |1-\lambda|: \lambda \in \sigma(L)\}$\\
ST~\cite{Grabow2010a}              & Synchronization time       
& $1/\max\left\{\sigma \left(\frac{L+L^\top}{2}\right) \setminus  \{0\} \right\}$
 \\
OSM               & Mean of out-shells       &
$\mathrm{mean}\left(\{ \mathrm{shell}^-( v)  \} \right)$    \\
OSV               & Variance of out-shells       &
$\mathrm{var}\left( \{ \mathrm{shell}^-( v) \} \right)$    \\
ISM               & Mean of in-shells       &
$\mathrm{mean}\left(\{ \mathrm{shell}^+( v) \} \right)$    \\
ISV               & Variance of in-shells       &
$\mathrm{var}\left( \{ \mathrm{shell}^+( v) 
 \} \right)$    \\
M               & Modularity       
& See~\cite{Leicht2008}  \\
\bottomrule \\
\end{tabular}
\end{center}
\end{table}

The descriptors were chosen such that many important aspects of complex networks
are sufficiently covered, while keeping computational effort manageable. 
They can be subdivided in three categories:
\begin{itemize}
\item degree statistics: we consider average of in- and out-statistics, their 
fluctuations and several type of correlations;
\item spectral statistics: we consider the spectral radius, average and fluctuations
of the eigenvalue spectrum and two different synchronization measures;
\item community structure: we consider average and fluctuations of the $k$-shell
statistics and of the clustering coefficient, as well as Newman's modularity.
\end{itemize}

We distinguish between ``local'' descriptors, which can be estimated by sampling small parts of the network, and ``global'' descriptors, for which knowledge of the full network is necessary. For example, to estimate the mean degree of the nodes in a network, it suffices to pick a number of  nodes one after an other and count their neighbors. However, the spectral radius of the connectivity matrix is not the sum of spectral radii of small parts of the network, but depends on the structure of the whole network and therefore cannot be estimated in this way.

We use the same symbol ($\mathrm{mean}$ or $\mathrm{var}$)
for both the theoretical value and its unbiased estimation.
Since the network parameters are independently chosen in every network realization, 
for fixed $c$, the numbers $f^{c_n}_i$ form a multivariate random variable whose realizations are independent over the instances $n$. 
As a consequence, dependencies between the $f^{c_n}_i$ originate from 
statistical links across features.

\subsection*{Feature extraction}
For computing the statistics in Figure 1-2 in we used 10000 networks 
with 100, 333 or 1000 nodes,
respectively, and with an overall connectivity of $p=0.1$.
For Figure 3 we used 4000 networks, where overall connectivity and node number
were matched with the corresponding statistics of the real networks.
We extracted the largest strongly connected component (LSCC) of each network
using a classical algorithm~\citep{Tarjan1972}.
All features were computed from the LSCC of the network.
Typically, the LSCC equaled the whole network for classical
network models or a large part of it in the case of MFs.
Networks with a largest connected components of a size smaller than 0.1 times the number of nodes were discarded.
Real datasets displayed different LSCC sizes:  274 (for 279 nodes, 2990 connections) 
for the \textit{C.\ elegans} neural network, 413 (456 nodes, 1014 connections) for the 
\textit{R.\ prowazekii} metabolic network and 904 (1022 nodes, 5075 connections) for the Roget synonym network.
After the calculation of network features, networks with undefined features were discarded. 
A typical case occurred for Watts-Strogatz networks with low rewiring: if the 
degree sequence is constant, its variance is 0 and many correlation
measures are undefined.
Nevertheless, this occurred only rarely (less than 5 networks in 1000 generated ones).

\subsection*{\ER networks}
These are the classical random networks~\citep{Erdos1959}.
Each connection is realized with probability $p$. 
Random networks of this type are, in fact, MF(1,1) networks.

It must be noted that \ER networks do not have any free parameter in our study,
since the connection probability is fixed. 
The only variability present in the \ER networks is due to the random realization
of the edges and not on the parameter choice.

\subsection*{Watts-Strogatz networks}

The Watt-Strogatz random network model~\citep{Watts1998} is constructed by
connecting nodes on a ring up to a certain geodesic distance. Then a rewiring
parameter $p_r$ is chosen and every edge is randomly rewired with a probability $p_r$.

We started with a ring network with a given number
of nodes. We then realized in-and out connections to $k=pN$ nearest neighbors 
such that the expected average degree is correct. 
Each connection is rewired to a randomly chosen target with a fixed rewiring probability $p_r$, randomly chosen for every network as a uniform random real between 0 and 1.

\subsection*{\BA networks}
Preferential attachment models like the \BA models prescribe that, as 
nodes are added to the network, their connections are drawn randomly
with a probability proportional to the degree of the target node.

For this study, we extend the classical preferential attachment model~\citep{Barabasi1999} in order to achieve a suitable randomization of statistics across networks. We also need to turn the graph into an oriented graph in such a way that the variances of the local features (across nodes) do not vanish.
As a first step, we drew a uniform random integer of nodes between the mean degree $D=pN$ and the desired number of nodes $N$.
Then, one node at the time was added, and bidirectional connections to 
existing nodes were established. Connection probability was proportional to the target degree, as in classical preferential attachment models.
This procedure continued until the number of nodes reached $N$.
Finally, we randomly break the network symmetry, by deleting every edge independently
with a certain probability, which was chosen in order to obtain the final desired
mean degree.

\subsection*{Equilibrium random networks}
Equilibrium random networks are characterized by a prescribed
expected degree sequence. Nodes are then connected to each other with a probability proportional to the product of their expected degrees.
This model has been recently introduced by Chung and Lu~\citep{Chung2002}.
A power-law degree sequence was generated with an exponent drawn uniformly between 0 and 2.

\subsection*{Multifractal networks}

The multifractal network generator has been recently introduced 
by Palla et al.~\citep{Palla2010,Palla2011}.
The basic idea is that networks are created from a generating measure $P$ on the unit square with 
a complex and variable structure, leading to very variable networks. 
Initially, we draw uniformly both a tuple of $n$ division lengths and a tuple of 
$n^2$ probabilities that sum up to $1$ each. The generating measure is then constructed in the following way: 
Initially, the unit square is divided 
into a $n^2$ rectangles by dividing the interval (0,1) on the $x$- and $y$-axes, respectively, into 
$n$ parts, according to the preset division lengths. The value of $P$ in each rectangle is assigned 
in turn according to the preset probabilities. In the next step, each rectangle is subdivided 
according to the preset division lengths, and the value of $P$ in each new rectangle is assigned from
the preset probabilities, multiplied by the value of $P$ of the current rectangle. Thus, each rectangle 
is replaced by a shrunk version of the whole generating measure of the previous step, times the value 
of $P$ in the current rectangle. This procedure is repeated for $k$ iterations, leading to an increasingly 
rough landscape, which for large $k$ approximates a singular defining measure~\citep{Lovasz2006}. \\
Once the generating measure $P$ has been produced, to obtain a network with $N$ nodes and a desired mean 
degree $k$, we replace $P$ by 
$P^\prime = \frac{k}{N} \frac{P(x,y)}{\int_0^1 \int_0^1 P(x,y)\,dx\,dy}$. \\
Each node $i$ is then given a position $x_i \in (0,1)$ and a connection from node $i$ to $j$ is made with a 
probability corresponding to $P(x_i,x_j)$. Deviating from the original proposal in~\citep{Palla2010}, 
we do not impose a symmetry condition on $P$ and draw each connection independently, to obtain directed 
networks. Parameters are randomized by choosing random tuples of divisions lengths and probabilities.

\section*{Results}



\begin{figure*}
\begin{center}
\includegraphics[width=\textwidth]{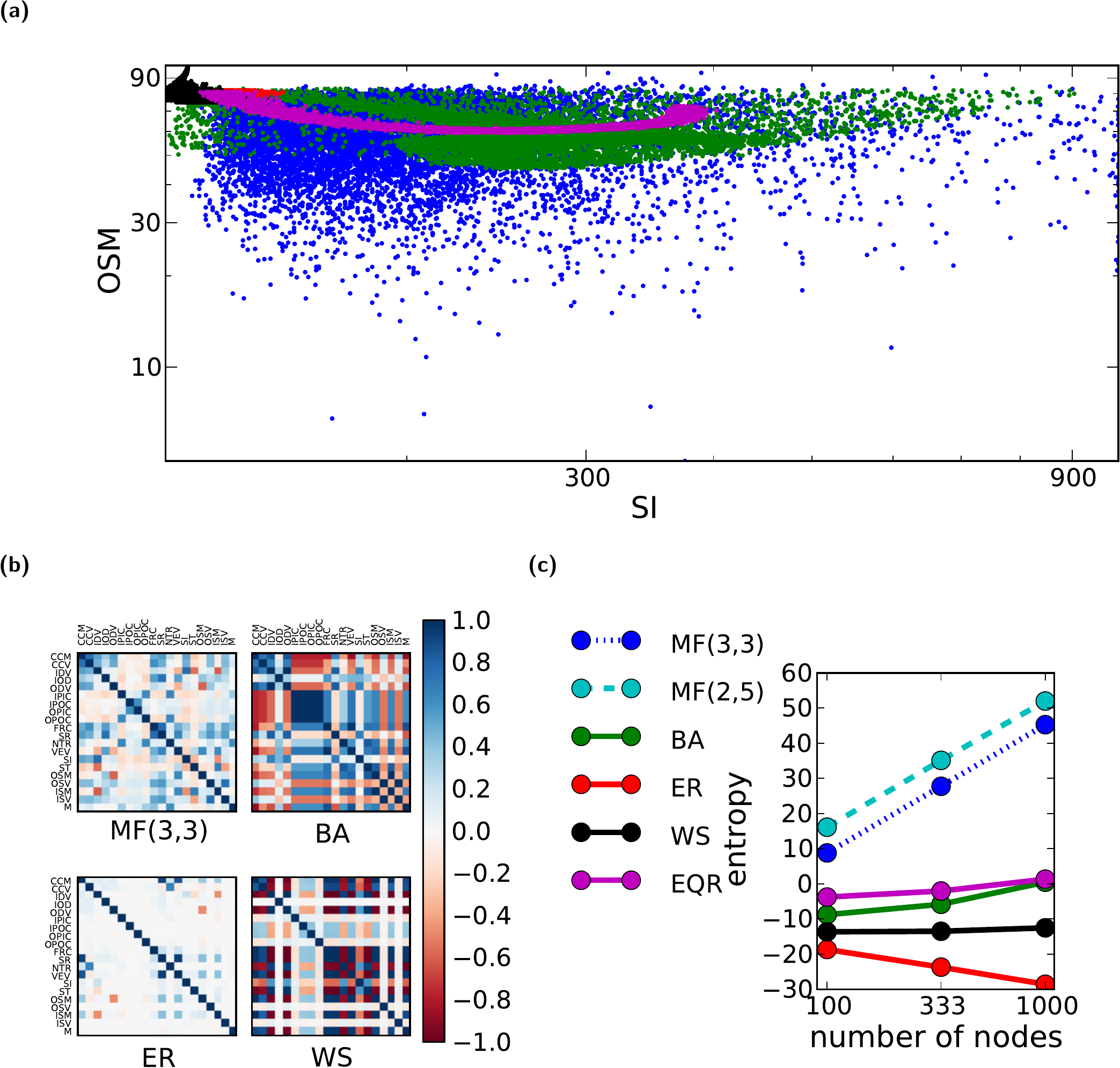}
\end{center}
\caption{{\bf Variability generated by various network models.}
(a) Scattered data of two global features for realizations of different types of networks (size $N=1000$), 
displayed in loglog scale. 
On the horizontal axis the synchronization
index SI, on the vertical axis the mean out $k$-shell OSM of the corresponding 
graph are shown. 
(b) Correlations between pairs of features, arranged in a matrix (size $N=1000$). 
For BA and WS networks, a clear 
structure is visible, due to the thematic ordering of the features.
Strong correlations are, in fact, the major cause for the low entropy generated 
by non-MF networks,
quantified in Panel (c).
Entropy of the multivariate distribution of features.
The entropy generated by MF networks is considerably higher, and it scales linearly
with the number of nodes in the networks.
}\label{fig:variability}
\end{figure*}

\subsection*{Variability of networks generated by different models}
Feature variability and dependencies between features vary significantly
between different network models.
Ideally, the specific construction principle of the network model should not introduce dependencies between independent features.
In fact, across our network samples, there are quite strong dependencies, as can be observed in Figure~\ref{fig:variability}, Panel (a). 
For several of the network models, scattered feature pairs for realizations of networks with 
random parameters are concentrated in a small, specific area of the 2-dimensional feature space.

These dependencies can be quantified by computing the matrix of pairwise correlation coefficients between features, computed across realizations of the same network model. ER and MF networks have apparently the least correlated features, whereas EQR, BA and WS networks have features with strong correlations.

However, not only the correlations between features determine the intrinsic
variability of a network model. The variability of the marginal distributions
must also be considered. 
To infer general laws of networks from samples, ideally, the network model should sample the complete space of network features in a uniform manner.  However, typically the sampled region of the feature space of a network model is bounded. The larger the variance of the features, the wider is the sampling of the model, and thus the greater the generalizability of the inferred statistical relations.
We estimate the overall variability $S$ of a given class of networks 
generated by our doubly stochastic process by the logarithm of the determinant of the covariance matrix $C$ of the features,
\[
S=\frac{1}{2}\log(\big(2\pi)^k\det(C)\big),
\]
where $k$ is the number of features.
For multivariate Gaussian distributions, this quantity corresponds to the Shannon entropy.  In a geometrical interpretation, $\det(C)$ represents a measure for the volume of the feature space the network model is able to sample. 

In Figure~\ref{fig:variability}, Panel (c) it is apparent that  
{the multifractal network generator (MF)} by far outperforms all other 
network models with regard to  feature variability.
It is interesting to note that {the variability of} WS
networks {is} considerably smaller than that {of} preferential attachment networks and 
equilibrium random networks.
This is an important issue to keep in mind, especially in view of the large number of studies inspired by 
the Watts-Strogatz network model~\citep{Wagner2001,Gerhard2011}.
{The generated entropy reflects only partially the number of 
degrees of freedom of the network models. On the one hand, since the overall connectivity
is fixed, ER networks do not have a single degree of freedom and they are the networks
with the least generated entropy, whereas MF networks generate the largest 
entropy, also thanks to their larger number of degrees of freedom. 
However, on a finer scale, the generated entropy also depends on other factors.
For example, the BA, EQR and WS network models all have one single degree of freedom,
but the latter performs considerably worse. Furthermore, MF(3,3) have 10 degrees
of freedom, 
but generate a lower entropy than MF(2,5), which only have 4 degrees of freedom.\\


\begin{figure*}
\begin{center}
\includegraphics[width=\textwidth]{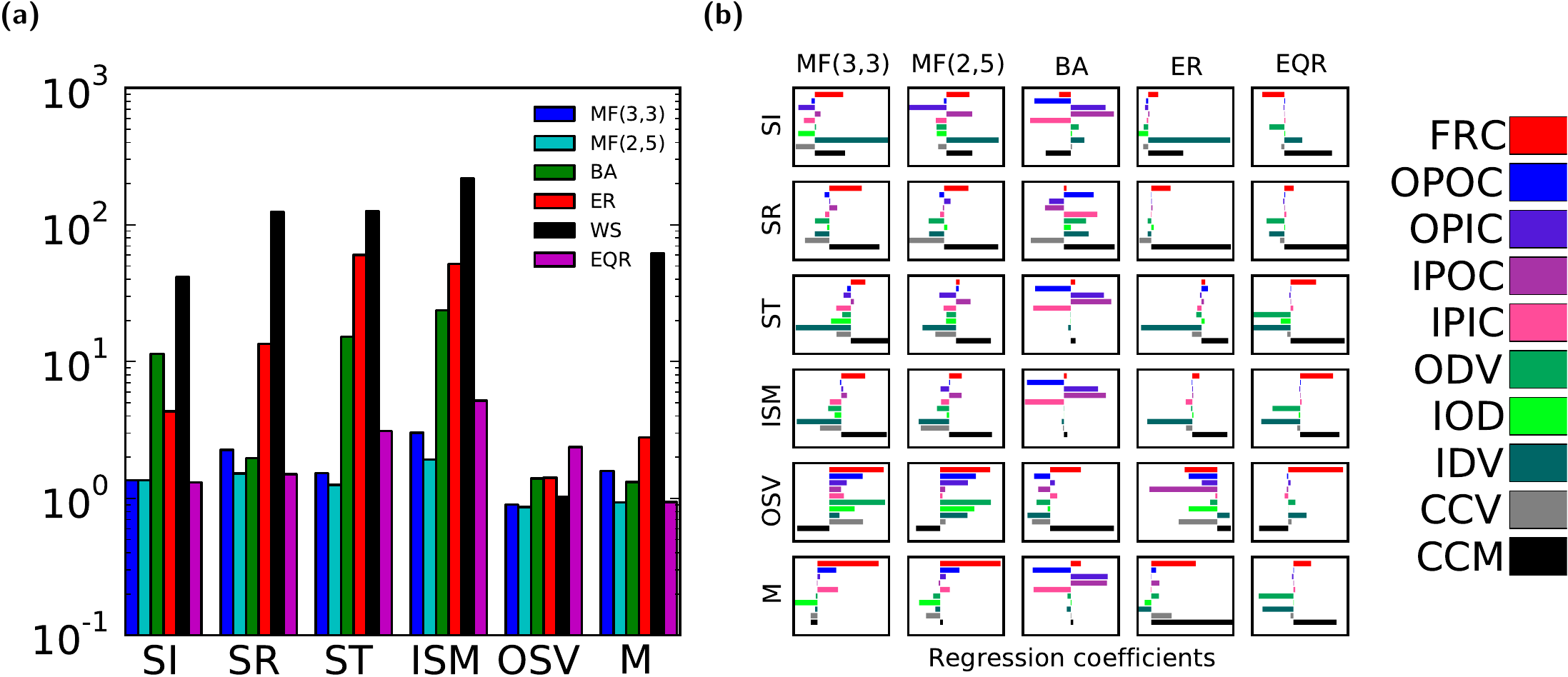}
\end{center}
\caption{{\bf Prediction of global features from local ones.}
(a) Residual prediction errors.
For the global features, we train a linear regression model with the data 
generated by one particular network model with random parameters 
and we test data from all models.
The residual prediction error is given by the mean-squared error normalized by
the overall standard deviation of the corresponding feature. 
A value of 1 indicates the result obtained if the true mean of the population 
was known and used as a predictor.
Note that using the empirical population mean as a predictor leads 
to a relative error larger than 1.
MF network models perform consistently around 1, whereas other models 
have occasionally very large errors.
(b) The coefficients of the linear regressor from the MF(3,3) set, normalized by the standard deviation of the local features used for the prediction. We excluded WS due to their very poor
performance here.
For some of the global features, the magnitude of the coefficients is consistent
over the network models. For example, the positive contribution 
of the variance of the in-degree
to the synchronization index and negative contribution to the 
synchronization time is consistent with the dynamic interpretation of these measures.
}\label{fig:prediction}
\end{figure*}

\subsection*{Predicting global features from local features}
It has repeatedly been pointed out~\citep{Motter2005,Kitsak2010} that local features of a network 
(e.g.\ degree distributions and degree correlations) are, when considered in isolation, 
not necessarily informative when it comes to predicting the dynamic properties 
of a network. 
On the other hand, global features (e.g.\ spectral properties and $k$-shell decomposition 
~\citep{Kitsak2010}) are difficult to obtain {for large networks} and, in general, 
are not robust against under-sampling of the network.

To overcome this problem, one could ask whether it is possible in principle 
to predict global features from a large set of simultaneously measured local ones. 
To test this idea, we trained for every network class a 
least-squares linear regressor on the vector of its local features to predict its global features. 
A distinct linear regressor was trained for every single global feature.
As a test set we used the full {data}set of networks of all classes. 
In Figure~\ref{fig:prediction}, Panel (a) 
we compare the performance of the different network models. To this end, a prediction for the global feature $x_i(a)$ of a realization $i$ of a certain network type $a$ was calculated using the local features of the specific realization and the linear regression coefficients obtained from networks of type $b$. As a measure for the deviations between these values and the predictions $\hat{x}_i(a,b)$ we consider the residual error
\[
\bar{\sigma}(x,a)=\frac{\Big[\frac{1}{AI}\sum_{a=1}^{A}\sum_{i=1}^{I}\big(x_i(a)-\hat{x}_i(a,b)\big)^2\Big]^{1/2}}{[\sum_{a=1}^A\sum_{i=1}^{I}x_i(a)]^{1/2}}
\]
where  $I$  indicates the total number of realizations of networks from each type and $A$ the number of network types. The normalization factor allows comparison between the performances for different features $x$.

Although least-squares linear regression is a rather simple approach to this complex problem,} this procedure allows to compare how well results from different network models can be generalized. Furthermore,
interesting information can be extracted from an examination 
of the regression coefficients, see Figure~\ref{fig:prediction}, Panel (b)
and our discussion below. 

\begin{figure*}[p]
\includegraphics[width=\textwidth]{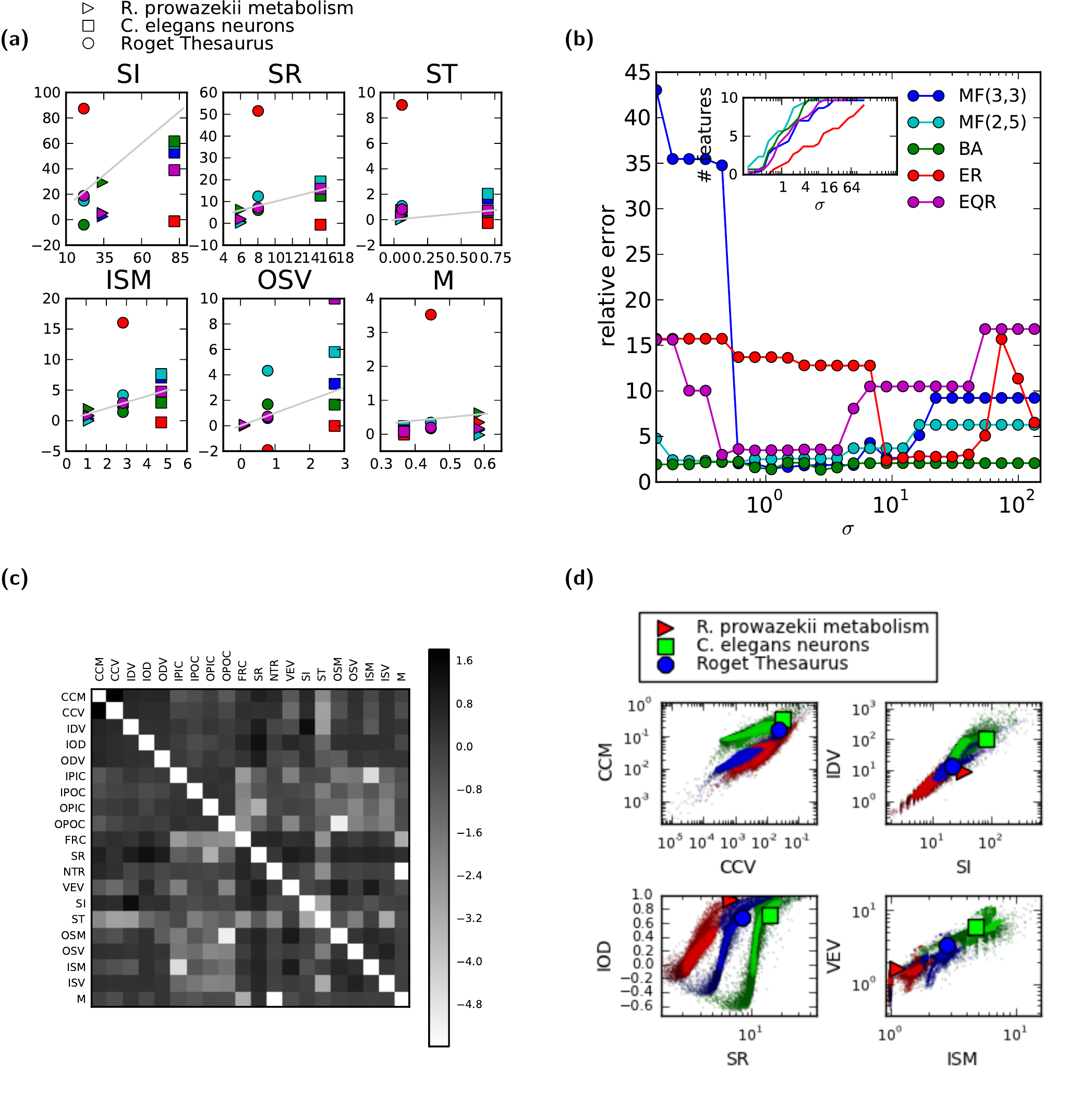}
\caption{{\bf Prediction of global features in real-world networks.}
(a) Scattered data of the predicted global features
for three data sets, using the regression coefficients obtained 
from network models with matched network size. 
Colors encode the model used for prediction.
(b) To study whether the prediction is robust with respect to the chosen threshold, 
we depict the relative mean-squared error 
(defined as in Figure~\ref{fig:prediction})
averaged over the whole data-set of real-world networks as it depends on the threshold.
The inset shows the average number of selected features for a given value of the
threshold $\sigma$.
(c) Reliability index 
$R(f_1,f_2)$ of the correlation coefficients
between pairs of features, calculated across network models. 
High values point toward a general statistical law for all networks.
(d) Data scatters for some pairs of features with significant correlations. 
Different colors encode different datasets: 
The number of nodes and the overall connectivity is extracted 
to generate a set of matched  networks from various models. 
The  scattered data are extracted from surrogate networks. 
The large markers denote the positions of the true dataset in the
data cloud. For pairs with a low CV across networks, 
the statistics of the real-world networks lie in the data cloud, suggesting that
those relations correspond to relevant statistical laws of complex networks.
In the upper left panel, the \emph{R.\ prowazekii} metabolism network is missing 
because of degenerate statistics. 
}\label{fig:reals} 
\end{figure*}

Finally, we studied whether {our} approach can be applied to real-world networks
extracted from publicly available datasets. 
We considered the connectome of
the nematode \emph{C.\ elegans}~\citep{Varshney2011b}, 
a synonym network based on the Roget's Thesaurus retrieved from the Pajek datasets collection~\citep{Pajek2006},
 and the metabolic network of the bacterium \emph{R.\, prowazekii}~\citep{Jeong2000}.  
Our selection was based on several criteria:
first, their size matched the size of the networks used for the evaluation of variability.
Furthermore, they represent directed {graphs} and have a large strongly connected component.
Finally, their physical/biological nature is quite diverse.
For each of the datasets we generated network ensembles as described above,
with matched number of nodes and average connectivity. 
On each network ensemble, we trained a linear regression model using an appropriate 
subset of  local features. 
The subset was chosen such that local features not represented well in the dataset
are excluded.
To this end, we fixed a threshold $\sigma$ and only used those local features 
the value of which did not deviate from the average value of the corresponding 
training set by more than $\sigma$ standard deviations. 
For each dataset we studied how the regression performance depends on the threshold. 
The performance was quantified by the relative mean-square error calculated 
over global features and networks, see Figure~\ref{fig:reals}.
For this purpose, all of the 
MF, EQR and BA networks resulted in regression models with quite good predictive power.

Furthermore, it is possible to use real networks as a cross-validation for
the statistical methods we are proposing.
To this aim, we first want to estimate the reliability of the correlation between
two features. This is done by computing a 2-dimensional matrix with the entries$$
R(f_1,f_2):=\log\left(\frac{
\left|\mathrm{mean}_g\left( CC_g(f_1,f_2) \right)\right|}
{\mathrm{var}_g(CC_g(f_1,f_2))}\right).
$$
Here $g$ varies over the network classes.
This matrix, depicted in Figure~\ref{fig:reals}, Panel (c), assesses the reliability
of a correlation between two features across models. The ten relations with the highest reliability index are listed in Table \ref{tab:relations}.

\begin{table}
\caption{Correlated feature pairs with highest reliability index}\label{tab:relations}
\begin{center}
\scriptsize
\begin{tabular}{ll}
\toprule
Feature 1  & Feature 2 \\
\midrule 
CCV & CCM \\
SI & IDV  \\
SR & IOD \\
OSM & VEV \\
ISM & VEV  \\
VEV & SR \\
SR & IDV \\
SR & ODV \\
SR & CCM \\
\bottomrule
\end{tabular}
\end{center}
\end{table}

To decide whether the relations between features are a peculiarity of 
the stochastic network models under consideration, we compare the model statistics with the true data previously introduced. 
If a relation between two features is of the same type both in real-world 
and model networks, then one would expect that the feature pair 
for the real-world network lies on the corresponding manifold for the
model networks.
Indeed, for the feature pairs with the highest $R$ values 
we verify in a scatter plot that the true data lie on the same manifold as the model data,
Figure~\ref{fig:reals}, Panel (d). 
We can thus conclude that a high $R(f_1,f_2)$ value is a good predictor
of the reliability of the correlation between a feature pair $f_1,f_2$.
This cross-validation method allowed us to reveal statistical laws for 
networks that would otherwise be quite difficult to discover. 
Three selected examples are highlighted below and,
in the following paragraph, we discuss  the synchronization properties
of networks in greater detail.
\begin{enumerate}
\item Mean and the variance of the clustering coefficient over the network
are consistently (positively) correlated across networks 
(mean Pearson's correlation $0.79$, standard deviation $0.12$).
As a consequence, properties 
attributed to the mean clustering coefficient~\citep{McGraw2005,Wu2008}
could be as well attributed to the variance of the clustering coefficient.
In this type of studies, additional considerations must be taken into account
to disentangle the contributions of these two measures.
\item The variance of the distribution of the eigenvalues 
(seen as a complex-valued random variable) is consistently (positively) correlated with
both the mean of the in- ($0.65\pm0.21$) and the out-$k$-shell decomposition 
($0.64\pm0.23$).
The mean in- and out $k$-shells encode,
roughly speaking, how well-connected the network is.
Local $k$-shell values are, as an example, predictive for epidemic spreading
efficiency~\citep{Kitsak2010}. We thus speculate about a role for eigenvalue
variance in determining the connectedness of a complex network.
Although this observation is purely heuristic, it could be of help
for scientists who use $k$-shell decompositions as a tool to understand the dynamics of 
complex networks.

\item The spectral radius is consistently (positively) correlated with
the mean clustering coefficient ($0.63\pm0.25$),
with the variance of the in-degree and the variance of the out-degree ($0.65\pm0.25$
in both cases), and with the in-out degree correlation ($0.72\pm0.2$).
The latter has an intuitive interpretation:
the spectral radius is related to the stability properties of an associated linear
system. The spectral radius $\rho(A)$ determines the asymptotic behavior
of the linear dynamical system defined by the recurrence equation $x_{n+1}=Ax_n$.
A high in-out degree correlation means that nodes receiving input from
many inputs project to many other nodes, thus destabilizing the system.
Finally, the spectral radius is, as expected, consistently (positively) correlated with
the eigenvalue variance ($0.70\pm0.27$).
\end{enumerate}


\subsection*{Synchronizability and in-degree variance}
The two features ``synchronization index'' and 
``in-degree variance'' are a very interesting case that deserves special attention. 
The synchronization index has been introduced
for directed graphs to quantify the degree to which a network is 
 prone to synchronization~\citep{Atay2006}. 
High values of this  index
correspond to  bad synchronizability of the network, while low values indicate 
that the networks synchronize easily.

For  MF, EQR and BA networks multivariate linear regression is most
efficient, and for these models 
the synchronization index and the in-degree variance have a correlation 
coefficient
of $0.85 \pm 0.04$. This is in marked contrast to the fact that these  networks are
of very different character: MF and EQR are locally of \ER type, whereas BA is not;
MF and BA networks typically have narrow unimodal SI distributions, whereas EQR networks
exhibit a peculiar uniform SI distribution. EQR and BA have a degree distribution with power-law tails, a property not shared by MF networks.

Our observations are in contrast to the conclusions previously drawn~\citep{Atay2006}
regarding the difficulty of predicting synchronizability by statistical
network properties. Our results imply that, 
{for real-world networks}, statistical properties can  indeed be informative
about spectral properties. We also have shown that {local} statistical
properties, as the variance of the in-degree, can be used to infer spectral
properties. It must be mentioned that related results have been analytically 
obtained for the case of non directed networks~\citep{Zhan2009}.These results shed  new light on the observation by Grabow et al.~\citep{Grabow2010a}
that networks in the small-world regime synchronize slowly. In fact, there
is a positive correlation $0.63 \pm 0.29$ between the variance of the
in-degree and the mean clustering coefficient, such that, in general, 
networks with high mean clustering coefficient have also a high variance of 
in-degrees and, therefore, synchronize slowly.

Furthermore, our results are perfectly consistent with recent results obtained in 
the theory of neuronal networks~\citep{Roxin2011a}. There, it has been shown that
in a model ensemble similar to our EQR setting, decreasing the variance of the
in-degree distribution leads to fast oscillations.

\section*{Discussion}
A significant amount of recent research has focused on non-random aspects of real biological networks,
especially in studies of metabolic interactions~\citep{Jeong2000}, 
of neuronal networks~\citep{Varshney2011b,Sporns2011,Prettejohn2011},
and of epidemic spreading~\citep{Kitsak2010}.
In neuroscience, in particular, the question has arisen of how different 
network features influence network performance with respect to different computational
aspects~\citep{Jarvis2010,Chen2011}.
In this type of works, different approaches have been used.
The first approach is to use data from related real-world data 
sets~\citep{Kötter2000,Kitsak2010,Rocha2011}.
One difficulty presented by this approach the generate surrogate data.
Degree preserving randomization has been suggested as a method for  assessing 
statistical significance of observed features in this approach~\citep{Milo2002,Artzy2004,Milo2004a}.

Alternatively, 
\emph{ad hoc} network models have been developed for studying the effect
of specific network features on the model dynamics~\citep{Albert2000,Boguna2008,Wu2008,MakiMarttunen2011}.
In this work, we assessed the generalization power 
offered by commonly used network models.
According to our analysis, a crucial limitation of most 
of the currently used network models is their low statistical variability 
in the  network features exhibited by the ensemble. 
This makes it unlikely that results obtained 
for a specific network model can be extrapolated to other contexts. 

In particular, the often employed WS (``small-world") model has quite singular statistical properties; on the one hand, 
the entropy generated by WS  networks with randomized wiring parameter is, at least for small networks, 
only slightly larger than the entropy generated by ER graphs, which have 
no free parameter when the mean connectivity is fixed.
In fact, ER networks are a special case of WS networks where the rewiring parameter is 1. On the other hand, 
WS graphs are outperformed by EQR networks with a randomized exponent of the degree distribution, which also 
have one degree of freedom, increasing the entropy of the ensemble.
It finally should be mentioned that the EQR model has some points in common with the degree preserving randomization algorithm proposed by 
Milo and coauthors~\citep{Milo2002}. 

{We found that the MF network generator~\citep{Palla2010,Palla2011} offers the possibility
to generate quite variable random networks with high predictive power. 
The entropy implied by these models is higher 
than the one generated by BA, EQR, WS and ER models. This property is due to 
the efficient use of a larger number of degrees of freedom in the network generating algorithm.
Moreover, in contrast to  other types of networks, the entropy of the ensemble seems to scale 
linearly  with the size of the networks in this case.
This property allows to reliably learn relations between local and global network features.}

Finally, and most importantly, we collected specific pieces of information 
regarding network properties by numerical experimentation.
A striking example concerns the negative correlation of the variance
of in-degrees with network synchronizability.
Results in this direction have already been obtained~\citep{Roxin2011a,Zhao2011}, although on
 specific topologies obtained with an algorithm similar to  EQR. Our results indicate that this
 may be a rather general property of dynamical systems on networks.
This finding could have important consequences, especially in view of the increasing evidence
 for a link between structural heterogeneity and stability in complex networks. 
Our method can be applied to include additional network features, like motif distributions,
 or characteristics of dynamical systems on networks,  and we would expect that 
further dependencies can be discovered between that have escaped our attention so far.


\acknowledgments{We thank Sadra Sadeh, Arvind Kumar and Jannis Vlachos for discussion and
valuable suggestions. Support by Bernd Wiebelt in setting up the cluster
computation is gratefully acknowledged.
Supported by the German Federal Ministry of Education and Research (BMBF; grant 01GQ0420 
"BCCN Freiburg", grant 01GQ0830 "BFNT Freiburg*T\"ubingen", grant 01GW0730 "Impulse Control"),
 and the German Research Foundation (DFG; CRC 780, project C4).
}


\bibliographystyle{apsrev4-1}
\bibliography{fractal_explorer}

\end{document}